\documentstyle[preprint,prl,aps]{revtex}
\begin{document}
\draft
\title{The Holstein Polaron}
\author{ Janez Bon\v ca$^a$, S. A. Trugman$^b$, and I.~Batisti\'c$^c$}
\address{$^a$FMF, University of Ljubljana and 
J. Stefan Institute, 1000, Slovenia,
$^b$Theory Division,
Los Alamos National Laboratory, Los Alamos, NM  87545,
$^c$Institute of Physics of the University, HR-1000, Zagreb, Croatia} 
\date{\today}
\maketitle
\begin{abstract}

We describe a variational method to solve the
Holstein model for an electron coupled to dynamical, quantum
phonons on an infinite lattice.
The variational space can be systematically expanded
to achieve high accuracy with modest computational
resources (12-digit accuracy for
the 1d polaron energy at intermediate coupling).
We compute ground and
low-lying excited state properties of the model 
at continuous values of the wavevector $k$
in essentially all parameter regimes.
Our results for the polaron energy band, effective
mass and correlation functions compare favorably with 
those of other numerical
techniques including DMRG, Global Local and exact diagonalization.
We find a phase transition for the first excited
state between a bound and unbound system of a polaron and an
additional phonon excitation.  The phase transition is also treated
in strong coupling perturbation theory.

\end{abstract}

\pacs{PACS:  71.38.+i, 72.10.Di, 71.35.Aa}

\section{Introduction}

Polaron formation as a consequence of electron-phonon
coupling appears in many contexts in condensed matter
physics, including charge carriers in colossal magnetoresistance
materials \cite{jaime,ramirez}
and in high-temperature superconductors \cite{bishop,alex}.
As theoretical research in this field spans over five decades, many analytical
techniques have been applied to this problem \cite{alex1}.  The
applicability of these methods is usually limited to a particular
parameter regime, frequently far from the physically most interesting
crossover regime.  Despite extensive analytical work in this field,
there remain many open problems including the nature of the
crossover to large polaron mass, 
the form of various correlation functions,
the nature of the polaron excited states, and dynamic properties of the
polaron.

Constantly growing computer capabilities have allowed 
research in this field to take advantage of various numerical methods,
including exact diagonalization techniques (ED)
\cite{alex2,wellein1,wellein2,wellein3,mello,capone,marsiglio},
quantum Monte Carlo calculations (QMC) \cite{raedt,kornilovitch},
variational methods including the Global-Local method (GL)
\cite{romero,romero1}, and recently developed density matrix renormalization
group techniques (DMRG) \cite{white}.  A comparison of
results obtained by different methods for
energy bands and effective masses is contained in the work of
Romero {\it et al.}  \cite{romero}.  Although most of these methods
give reliable results in a wide range of parameter regimes, each
suffers from different shortcomings. The ED technique gives reliable
results on small lattices (up to 20 sites \cite{wellein2}) for
ground and excited state properties. Results are limited to discrete 
momentum $k-$points. QMC methods can treat large system sizes (over 1000 lattice
sites) and provide accurate results for thermodynamic polaron
properties.  Dynamic properties, however, require analytic continuation
from imaginary time, which is an ill-posed problem that is extremely
sensitive to statistical noise.  The GL method gives reliable results
for energy bands and the polaron effective mass on reasonably large system
sizes (32 sites), however, it is limited to
ground state properties. The DMRG method seems to be most successful in
computing ground state properties including various correlation
functions. Despite the lack of translational symmetry, it
provides reasonably accurate results for energy bands and the
effective mass. It can deal with large system sizes ({\it e.g.} 80
sites and 30 phonons per site), delivering  reliable results in the
intermediate and strong coupling regime. DMRG is less reliable in the weak
coupling regime.  It generally treats large systems with open
boundary conditions and does not allow the calculation of dynamic
or large $k$ properties or excited states, although there
are some exceptions to these rules \cite{jeckelmann}.

In this paper we focus on several issues of the Holstein problem.  We
present a simple and computationally efficient method based on the
exact diagonalization technique. We apply the method to the
one-dimensional, single-electron Holstein model. Among the advantages
of our method are the simplicity of our approach, the efficiency in
selecting the variational space, and the ability to compute 
both ground and excited state physical
properties of the system at continuous total wavevector $k$. 
We define
the variational space on an infinite lattice.  Even though most of our
calculations were done on a workstation, our results 
are often superior to those of
other numerically more intensive methods
\cite{wellein2,wellein3,white}.  We test the method by comparing
results for the energy band, effective mass, quasiparticle weight
and correlation functions with some of the most successful recent
numerical methods. 

The second part of this paper is devoted to the investigation of the
first excited state of the model. Using our numerical method as
well as a strong coupling analytical approach, we find a phase
transition between a state where the polaron forms
a bound state with an additional phonon excitation,
and one where there is
no bound state.

Generalizations of this method can even be used to calculate
coherent quantum dynamics far from equilibrium, including those of a polaron
driven by a strong electric field \cite{janez1}, and of an
electron tunneling across a potential drop while
coupled to phonons \cite{janez2}.

We consider the  Holstein Hamiltonian  \cite{mahan}

\begin{eqnarray}
H && = -t \sum_{j} ( c_{j}^\dagger c_{j+1} + H.c.) \nonumber \\
&& -\lambda \sum_j c_{j}^\dagger c_{j}
( a_j + a_j^\dagger) + \omega \sum_j a_j^\dagger a_j ,
\label{ham}
\end{eqnarray}
where $c_{j}^\dagger$ creates an electron and
$a_{j}^\dagger$ creates a phonon on site $j$.  We consider the case
where a single electron hops between nearest neighbor Wannier orbitals in 1d,
and interacts with dispersionless optical phonons.  The electron-phonon
coupling strength is $\lambda$, and is local in real space
($k$-independent).  The parameters $t$, $\omega$, and $\lambda$
all have units of energy, and can be used to form
two dimensionless ratios.  (Different conventions for the parameters
are sometimes used in other papers.)

It is clear at the outset that for any finite values of the
parameters, the exact groundstate will be a delocalized state with
momentum $\vec k$, and not a ``self-trapped'' solution that breaks
translation-invariance.  The simple argument is that if the ground
state were self-trapped, its center could be shifted and a degenerate
ground state obtained.  If the Hamiltonian has any nonzero matrix
element between states with different centers, a lower energy state
can be obtained by a phased superposition of wavefunctions with
different centers (a Bloch state).  The only known way that this
argument can fail is with (unphysically) strong electron-phonon
coupling as $\omega \rightarrow 0$
to a gapless phonon spectrum, 
in which case the matrix element
between different centers can vanish \cite{leggett}.
In that case, an exact ground state can be written as
either self-trapped or as a Bloch state.  Self-trapping cannot occur
here because the phonon spectrum has a gap.  Although the exact
ground state is a Bloch state, for some parameters (e.g.~very large
$\lambda$), a self-trapped solution can have a low variational
energy\cite{holstein}.  
It has indeed been proven that polaron ground state properties, 
including the
energy and effective mass, are analytic functions of the Hamiltonian
parameters \cite{gerlach}.  We will see below that such a theorem
cannot be extended to include excited states.

\section{The method}

A complete set of basis states
for the many-body Hilbert space can be written
\begin{equation}
\vert M \rangle =
  |j; \dots ,n_{j-1}, n_{j} , n_{j+1},  \dots \rangle,
\label{psi}
\end{equation}
where $j$ is the electron site and there are $n_m$ phonons on site $m$.
A variational subspace is
constructed beginning with an initial state, taken to be an
electron on site $j=0$ with no phonons,
and operating repeatedly ($N_h$-times) with the
off-diagonal pieces of the Hamiltonian, Eq.~(\ref{ham}).  
At each step, a basis state
is added when there is a nonzero $t$ or $\lambda$ matrix element to a
state previously in the space.  These states and all
of their translations on an infinite lattice are included
in the variational space.  (A translation moves the electron
and all phonon excitations $m$ sites to the right.)
If a basis state can be generated in more than one way,
only one copy is retained.  All nonzero matrix elements
of the Hamiltonian between retained basis states are included.

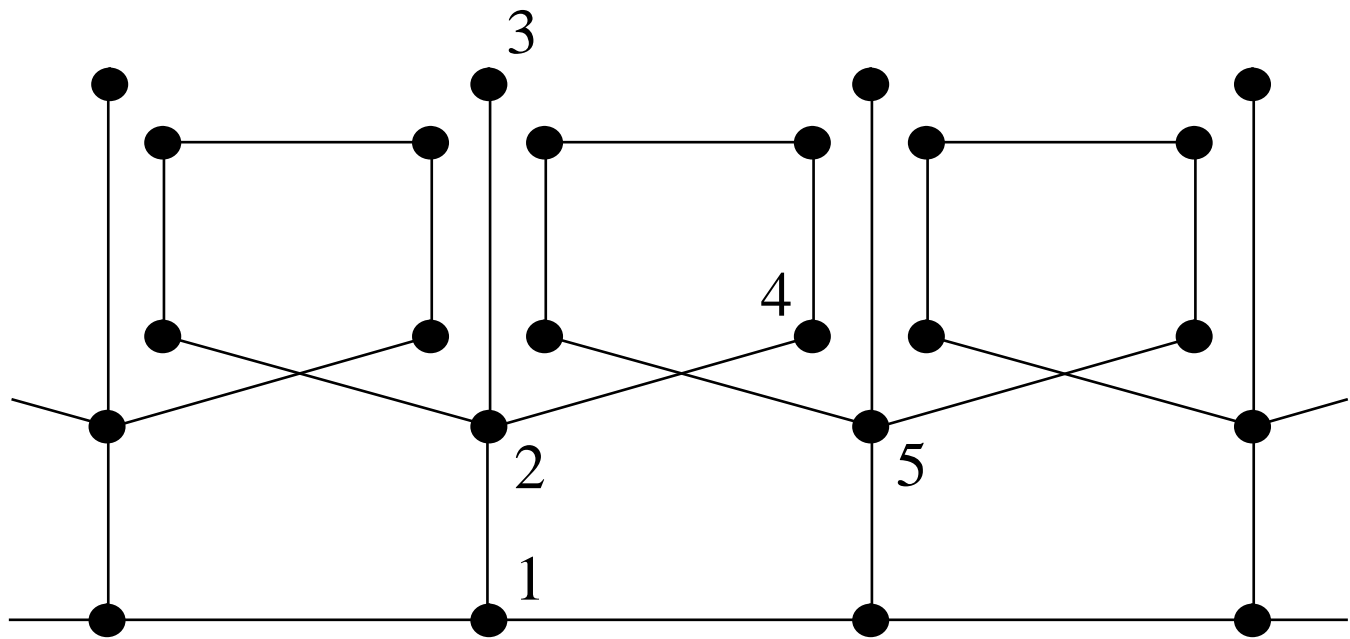
\begin{figure}
\caption[]{The small variational Hilbert space shown for the polaron is 
a subset of the $N_h = 3$ space.  Basis states in the
many-body Hilbert space are
represented by dots, and nonzero off-diagonal matrix elements
by lines.  The x-coordinate of the dots is
(aside from small displacements) the coordinate of the electron.
Vertical bonds create phonons, and horizontal
or nearly horizontal
bonds are electron hops.
State $|1 \rangle$ is an electron on site 0
and no phonons.  
State $|2 \rangle$ is an electron and phonon, both on site 0.  
State $|4 \rangle$ is an electron on site 1 and a phonon on site 0,
which is reached from state $|2 \rangle$ by hopping the electron
to the right.
State $|5 \rangle$ is a translation of state $|2 \rangle$.
The Hamiltonian is sparse in this basis, with at most
4 bonds attached to a dot.
The dots can also be thought of as
Wannier orbitals in a one-body periodic tight-binding model.
}
\label{tight}
\end{figure}

A small variational Hilbert space is shown in Fig.~(\ref{tight}).
The dots may be thought of as basis states in the
many-body Hilbert space, or alternatively as 
Wannier orbitals in a periodic (one-body) tight-binding model,
with hopping matrix elements given by the bonds.
The variational space shown in Fig.~(\ref{tight})
is still infinite, since the lattice repeats periodically
to infinity.  It is clear from Bloch's theorem, however,
that each eigenstate can be written as ${\rm e}^{ikj} \Psi _m $,
where $k$ is the momentum, $j$ is the unit cell, and $\Psi _m $ is a set of
$M$ complex amplitudes, one for each state
in the unit cell ($M=7$ for the example of Fig.~\ref{tight}).
For a given momentum $k$, the
resulting numerical problem is to find the eigenstates of an 
$M \times M$ (sparse) hermitian matrix using Lanczos 
or another method.

Figure (\ref{spaghetti}) plots the energy eigenvalues
for a larger variational space
containing a maximum of 9 phonon excitations.
The figure superficially resembles a ``band structure'',
which however encodes ground and excited state information for
the many-body (many phonon) polaron problem.  
The AC conductivity of the polaron,
for example, appears as an ``interband'' transition in
this mapping.

The largest variational basis that we have used
has $N_h=22$, or $M=1.2 \times 10^7$
states.  It is usually unnecessary to use such a large
basis for intermediate-coupling
ground state properties, even to obtain 12-digit
accuracy for the energy.   
The variational Hilbert space we construct is not 
a standard one,
and appears to add basis states more efficiently
than some other methods.  
A basis state is included if it can be reached using
$N_ \lambda$ phonon creation operators and
$N_t$ electron hops in any order
with $N_t + N_ \lambda \leq N_h$.
For a given  $N_h$, there is
a basis state with $N_h$ phonon quanta on the same site
as the electron and no phonon excitations elsewhere.  
There is also a basis state with $N_h - 1$ quanta
on the site adjacent to the electron and no phonon excitations elsewhere.
For $N_h$ odd, there is a basis state with $(N_h -1)/2 $
quanta on the electron site and an equal number simultaneously on
a first neighbor site.
The maximum distance that a phonon excitation can appear
from the electron is $l = N_h - 1$, but then only
a single quantum and only if there are no phonons excited
elsewhere in the system \cite{trade}.  Only the basis states
from a single unit cell (a single electron position)
are stored in computer memory.

The energy of a polaron for a finite chain of $N$ sites with periodic 
boundary conditions is {\it lower} than that
for an infinite lattice with the
same parameters.  (This is easily to verify
in weak-coupling perturbation theory, or numerically.)
Thus, previous exact diagonalization and most other variational approaches
produce energies that are variational for the particular
lattice size $N$ that they treat, but are not variational
in the thermodynamic limit $N \rightarrow \infty$.
The energy we calculate is, in contrast, variational
in the thermodynamic limit.
(The quoted DMRG energies are extrapolated and not variational, although they
are generally rather accurate.)

Having the capability to compute the polaron energy $E(k)$
at any $k$ rather than being limited to multiples of
$2 \pi / N$ makes our method more
accurate for computing the effective mass of the polaron using the
standard formula
\begin{equation}
  {m_0 \over m^*} =  {1 \over {2t}} 
  {\partial ^ 2 E(k) \over {\partial k ^ 2}}\vert_{k=0},
\label{mass}
\end{equation}
where $m_0 = 1/(2t)$ is the effective mass of a free electron.
The second derivative is evaluated by small finite differences in the
neighborhood of $k = 0 $.  Note that although the calculated
energy $E(k)$ is a variational bound for the exact energy, there is no
such control on the mass, which may be either above or below the exact
answer, and is expected to be more difficult to obtain accurately.
Nevertheless, in the intermediate coupling regime where our method at
$N_h=20$ gives an energy accuracy of 12 decimal places, one can
calculate the effective mass extremely accurately (6-8 decimal places)
by letting $\Delta k \to 0$.

Further information about the quasiparticle may be obtained by
computing the quasiparticle residue,
the overlap (squared) between a bare electron and a polaron,
\begin{equation}
Z_k = \vert\langle\psi_k\vert c_k^\dagger\vert0\rangle\vert^2,
\label{zk}
\end{equation}
where $\vert0\rangle$ is the state with no electron and
no phonon excitations,
and $\vert \psi_k\rangle$ is the polaron wavefunction
at momentum $k$.  The numerical results for $Z_{k=0}$ 
given in the next section differ by less than 1\% from results for
the reciprocal effective mass
$m_0/m^*$ obtained from Eq.~(\ref{mass}).  At finite $k$, $Z_k$ provides
information about the electronic character of the polaronic state. The
phonon contribution to the quasiparticle can be measured by the 
$k-$dependent mean phonon number
\begin{equation}
N^{ph}_k=\sum_i  \vert\langle\psi_k\vert a_i^\dagger a_i\vert\psi_k
\rangle\vert^2.
\label{Nk}
\end{equation}

To describe the polaron
at different $k$ we have also computed the static correlation function
between the electron position and the oscillator displacement
\begin{equation}
\chi(i-j)  = \langle\psi_k\vert c_i^\dagger c_i (a_j + a^\dagger_j)
\vert\psi_k\rangle,
\label{chi_eq}
\end{equation}
and the  distribution of the number of excited phonons in the vicinity
of the electron
\begin{equation}
\gamma(i-j)  = \langle\psi_k\vert c_i^\dagger c_ia^\dagger_ja_j
\vert\psi_k\rangle.
\label{gamma}
\end{equation}

\section{Ground state results}

In this section, we compare our results for the 
ground state properties of the Holstein model
to those obtained by other numerical methods.
We also calculate polaron correlation functions
at finite values of the Bloch wavevector $k$.

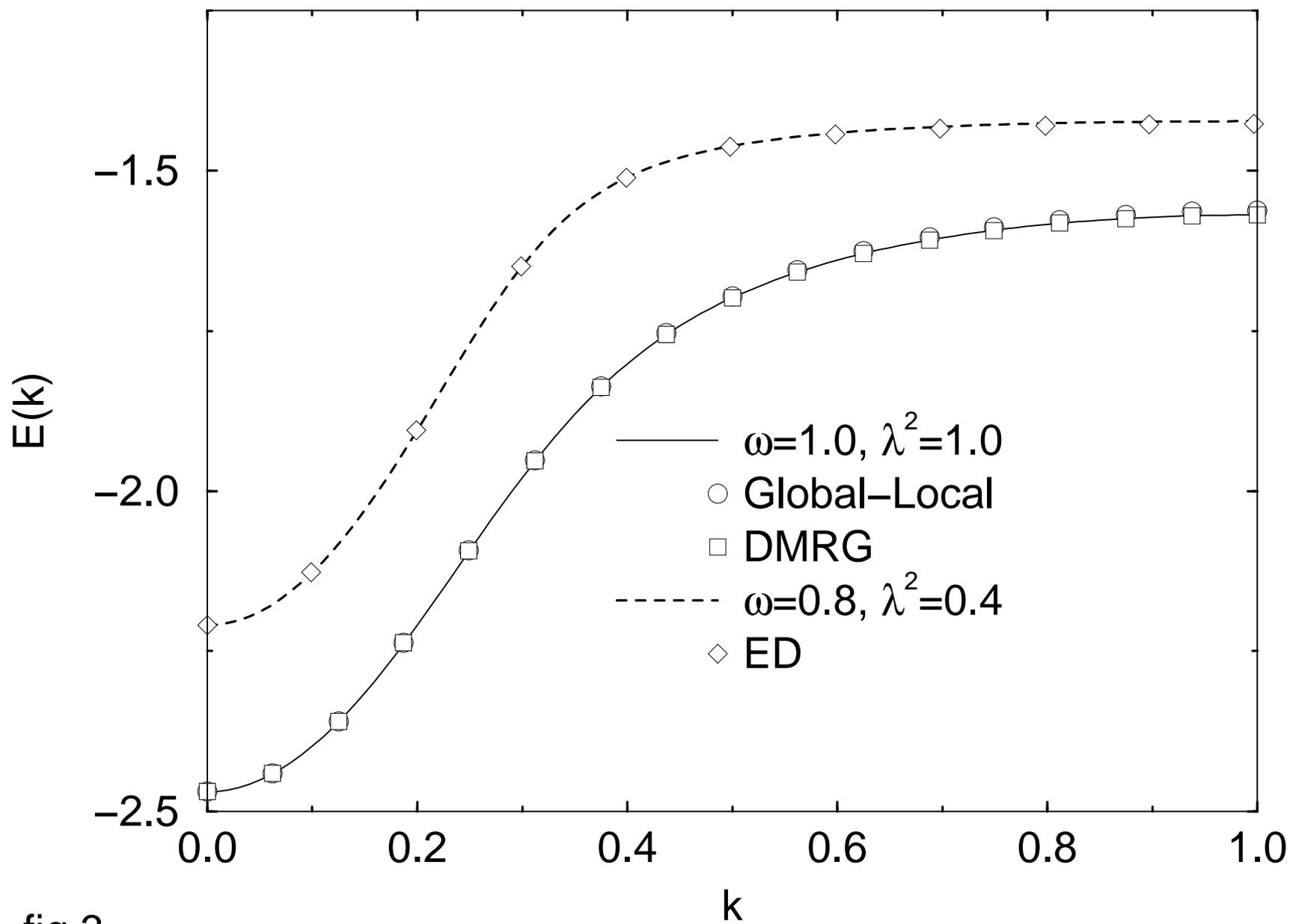
\begin{figure}
\caption[] {Polaron energy as a function of $k$ (in units of $\pi$).
Lines represent our results for two different sets of parameters
obtained by calculating $E(k)$ at
100 k-points.  Circles, squares and diamonds are results obtained with
Global-Local \cite{romero}, DMRG \cite{romero,white} and ED
\cite{wellein2} calculations respectively.}
\label{ekfig}
\end{figure}

We start by comparing energy bands $E(k)$ for two different sets of
parameters.  Figure (\ref{ekfig}) compares the present method to the
DMRG \cite{romero,white}, GL \cite{romero} and the finite cluster ED technique
\cite{wellein2}.   
Using $N_h=20$ we achieve an accuracy in
the thermodynamic limit of 12 decimal places for small $k$ and at least 4
decimal places  
for large $k$.  Our results for $E(k)$ are
presented as continuous curves. Agreement of our results with DMRG
and ED is good, however there is a slight disagreement with the
Global-Local method at larger $k$, which also disagrees with the DMRG
method. There is also a slight disagreement with the ED at larger $k$
which we attribute to the smaller system size used in finite cluster ED
calculations.  As we will demonstrate later in this paper, the extent
of the lattice deformation (the size of the polaron) increases as $k$
approaches the Brillouin-zone boundary, which makes finite-size
calculations more susceptible to finite size effects.

In table (\ref{energ}) we compare the polaron ground state energy
at $k=0$ for two different parameter sets obtained by
our and three other numerical methods, 
exact diagonalization (ED) \cite{fehske},
DMRG \cite{white}, and
Global-Local variational \cite{romero}.
Comparisons with a greater variety of
methods can be found in Romero {\it et al.} \cite{romero}.
We have limited our comparison to those methods that are accurate to
at least 4 decimal places.  Our method 
converges to all of the digits shown using $N_h=15$ or $M=88052$
basis states for $\lambda = \omega=1$,
and $N_h=18$ or $M=731027$ basis
states for $\lambda=\sqrt{2}$, $\omega=1$.  
The ground state energy appears to converge exponentially with
$N_h$, with the accuracy improving by approximately one order
of magnitude as $N_h$ is increased by 1.
The larger ($N_h=18$) calculation runs in under a minute
on a modest workstation.

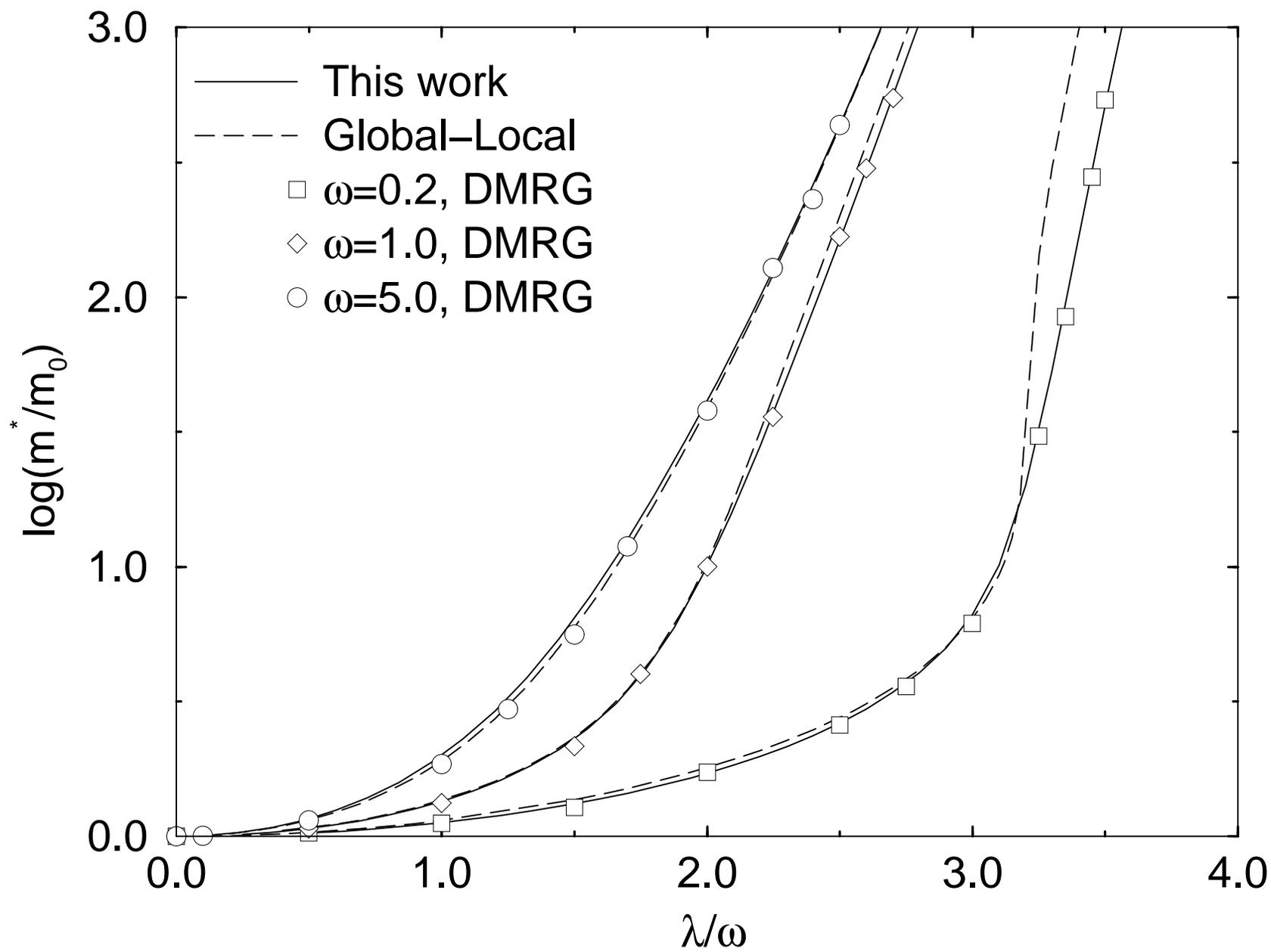
\begin{figure}
\caption[] {The logarithm of the effective mass $m^*/m_0$ as a function
of $\lambda/\omega$.  Our results are plotted as full lines,
and Global-Local results as dashed lines \cite{romero}.
Open symbols, indicating the value of $\omega$,
are DMRG results \cite{white}.
}
\label{mass0}
\end{figure}

Figure (\ref{mass0}) shows our results for the effective mass
Eq.~(\ref{mass}) computed with $N_h=20$ in comparison with GL and DMRG
methods.  The parameters span different physical regimes including
weak and strong coupling (respectively small and large
$\lambda/\omega$), and adiabatic ($\omega/t \ll 1$) and antiadiabatic
($\omega/t \gg 1$) regimes.  We find good agreement with GL away from
strong coupling and good agreement in all regimes with DMRG.  DMRG
calculations are not based on finite-$k$ calculations due to a
lack of periodic boundary conditions, so they
extrapolate the effective mass from the ground state data using chains
of different sizes, which leads to larger error bars and more
computational effort. Notice that their discrete data is slightly
scattered around our curves.  Nevertheless, both methods agree well. A
closer look at effective masses in the antiadiabatic regime
($\omega/t=5.0$) and $\lambda/\omega<2 $ reveals a slight systematic
disagreement of our results with both methods in comparison.  We have
compared our results for effective mass obtained on different systems
from $N_h=16$ with $M= 178617$ states to $N_h=20$ with $M=2975104$
states and obtained convergence of results to at least 4 decimal
places in all parameter regimes presented in Fig.~(\ref{mass0}). 
Our error is therefore well below the linewidth. Even
though there is no phase transition in the ground state of the model,
the polaron becomes extremely heavy in the strong coupling
regime.  The crossover to a regime of large polaron mass is more rapid in
adiabatic regime (smaller $\omega/t$).

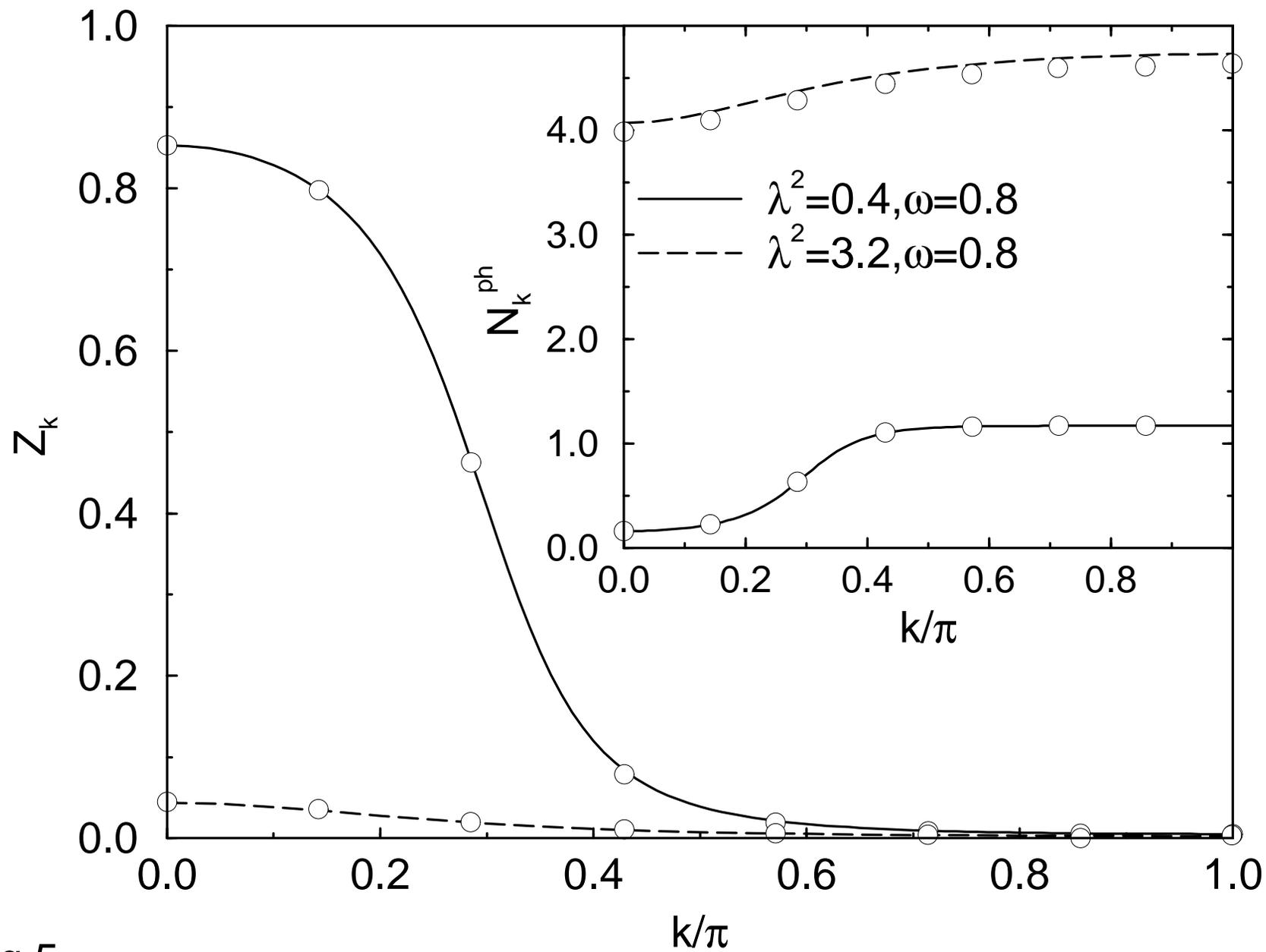
\begin{figure}
\caption[] {The quasiparticle weight $Z_k$ and in the inset 
the total number of 
phonons $N^{ph}_k$ as a function of the wavevector $k$. 
Results by Wellein and Fehske \cite{wellein2} are represented by open circles.}
\label{zkfig}
\end{figure}

Figure (\ref{zkfig}) shows the quasiparticle residue
$Z_k$ and the mean phonon number $N_k^{ph}$ 
as a function of $k$ for the case of small 
$\lambda_0 \equiv \lambda^2/2\omega t $ ($\lambda^2=0.4,  \omega=0.8$)
and large $\lambda_0$ ($\lambda^2=3.2, \omega=0.8$).
The two sets of parameters  correspond to the large and small polaron 
regime respectively \cite{wellein3}.
The DMRG cannot straightforwardly compute this quantity,
and we compare our
results with cluster calculations.  Open symbols represent the
results of Wellein and Fehske \cite{wellein2} obtained on a $N=14$ site
cluster for the same choice of parameters.  Except for the fact that
their results are limited to discrete $k-$points defined by the size
of their system, we find excellent agreement
between the two methods in the weak coupling case.
In the strong coupling regime there
is an approximately $1\%$ disagreement in $N^{ph}_k$ due to a lack of 
phonon degrees of freedom in the variational space of the
ED calculation.  
Our results in the weak coupling case show a smooth crossover
from predominantly electronic character of the wavefunction 
for small $k$ (large $Z_k$ and small $ N^{ph}_k \approx 0$)
to predominantly phonon character around $k =\pi$ characterized by
$Z_k\approx 0$ and $N^{ph}_k \approx 1$. In the strong coupling regime 
there is less $k-$dependence.
The $Z_k$ is already close to zero at small $k$, indicating strong
electron-phonon interactions that lead to a large polaron mass.

Jeckelmann and White have calculated the electron-phonon correlations
at $k=0$ for the 1d and 2d polaron using the DMRG
method \cite{white}.  We compare our results with theirs for the 1d case.
There is good agreement in Fig.~(\ref{chi}a) for intermediate
coupling parameters $\omega = 1$ and $\lambda=0.5$.
Figure (\ref{chi}b) plots correlations for $\omega=0.1$ and $\lambda=0.1$,
which corresponds to weak coupling in the adiabatic limit.
In this regime the DMRG method gives less reliable results.  The size of
the polaron is underestimated, possibly due to finite-size effects
in open boundary conditions.
Note also that the DMRG method does not give symmetric results
as it should, {\it i.e.}
$\chi(l)\not = \chi(-l)$.
In this parameter regime our results have 
fully converged, as we can see from the perfect overlap of results of
systems with two different sizes of the Hilbert space
$N_h=17,18$. 
Figure (\ref{chi}c) plots correlations for
$\omega = 0.1$ and $\lambda=0.435$, which belongs to the strong coupling,
small polaron regime. 
The DMRG produces superior results
in this regime, where our calculation at $N_h=21$ has not fully
converged to the large $N_h$ limit.  Our results are, nevertheless,
in qualitative agreement with the DMRG.
We conclude
that both techniques give reliable results in the intermediate
coupling regime, and that they complement each other in the weak and strong
coupling regimes. 

A thorough investigation of correlation
functions using ED and the variational Lanczos method was performed by
Wellein and Fehske \cite{wellein3}. Although we do not show a
direct comparison with their work, our
results for correlation functions agree qualitatively with their calculations.

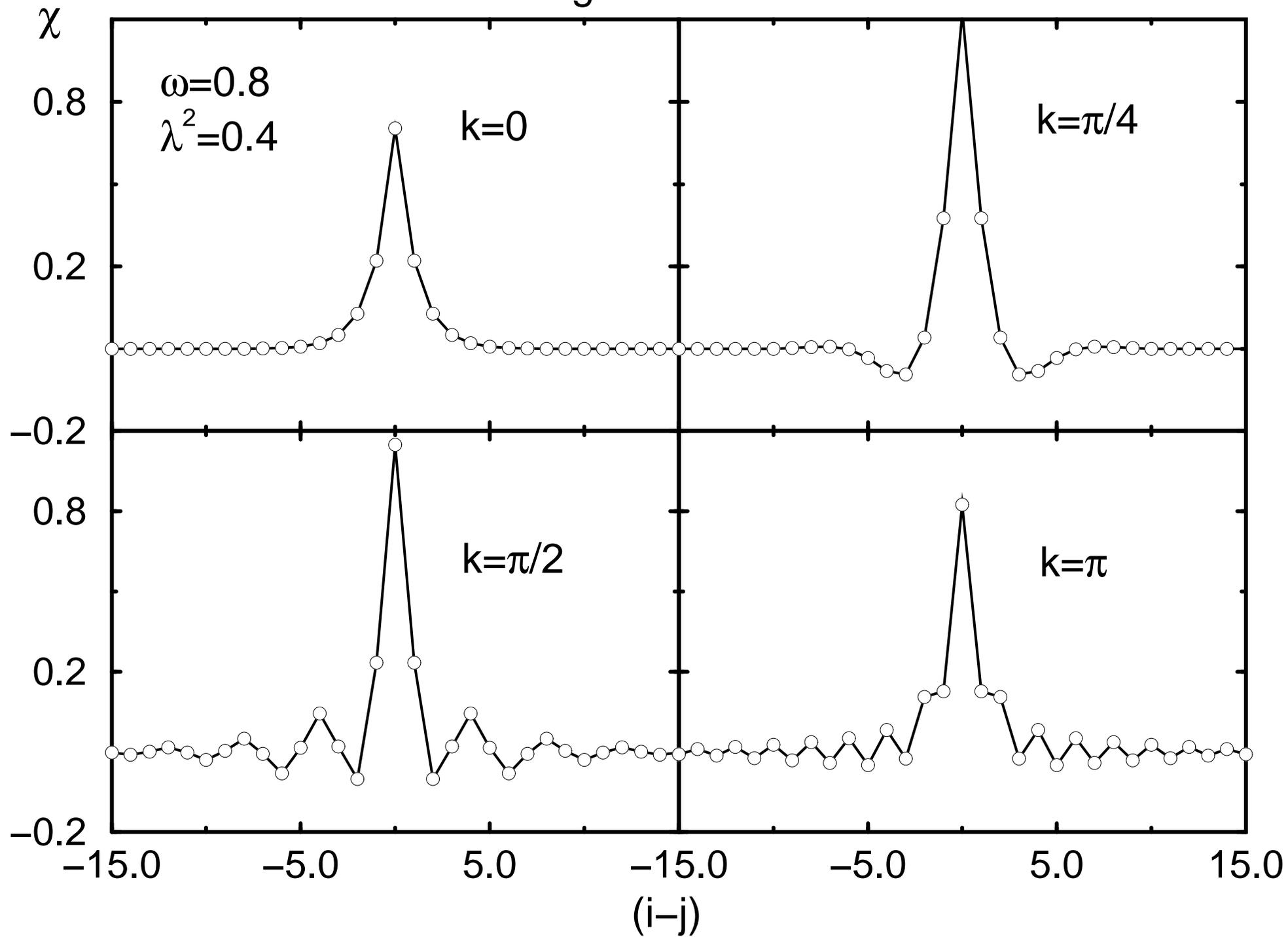
\begin{figure}
\caption[] {Lattice deformation $\chi$ as a function  of $(i-j)$ for
$\omega =0.8$, $\lambda^2=0.4$ and $N_h=18$, for four different values of
momentum $k$.  The variational Hilbert space for
$N_h=18$ allows nonzero correlations to a maximum distance
$\vert i-j\vert_{max}=17$.  Only distances up to 15 are plotted.
} 
\label{chifig}
\end{figure}

While the strength of the DMRG calculation is exhibited in its ability to
compute ground state properties of large systems, it is limited in its
computation of excited states.  In Figure (\ref{zkfig}) we have
shown how the nature of the polaron transforms from predominantly
electronic character at $k=0$ to phononic around $k=\pi$.  We 
follow this transformation by computing the correlation
function $\chi$ for four different
values of $k$, shown in Figure (\ref{chifig}). 
These parameters correspond to the weak coupling case in
Figure (\ref{zkfig}).  At $k=0$, where the group velocity is zero,
the deformation is limited to only a few
lattice sites around the electron.  It is always positive and
exponentially decaying.  At finite but small $k=\pi/4$, the
deformation around the electron increases in amplitude 
and rings (oscillates in sign) as the polaron acquires
a finite group velocity.
At $k=\pi/2$ the ringing is strongly
enhanced.  Note also that the spatial extent of the deformation
increases in comparison to $k=0$. The range of the deformation is
maximum at $k=\pi$, where it extends over the entire region
shown in the figure.  In keeping with the larger extent of the lattice
deformation near $k=\pi$, the ground state energy $E(\pi )$ converges
more slowly with the size of the Hilbert space.

We have also computed $\chi$ for the strong-coupling case $\omega=0.8,
\lambda^2=3.2$ (not shown).  We find only  weak  $k-$dependence, which is a
consequence of the crossover to the small polaron regime.  The
lattice deformation is localized predominantly on the electron site.

\section{What is the nature of the first excited state?}

In this section, we
focus on the question of whether an extra phonon excitation 
forms a bound state with the polaron,
or instead remain as two widely separated entities. 
Using numerical and analytical approaches we will show that there
exists a sharp phase transition between these two states.
Although the ground state energy $E_0$ is an analytic
function of the parameters in the Hamiltonian,
there are points at which the energy $E_1$
of the first excited state is nonanalytic.
In previous work, Gogolin has found bound states of the
polaron and additional phonon(s), but he does not
obtain a phase transition between bound and unbound
states because his approximations are limited to
strong coupling $\lambda / \omega \gg 1$ \cite{gogolin}.
A phase transition between a bound and unbound first
excited state has been calculated for dimension $d=3$
using a dynamical CPA approximation \cite{sumi} and dynamical
mean field theory \cite{ciuchi}.

\subsection{Numerical results}

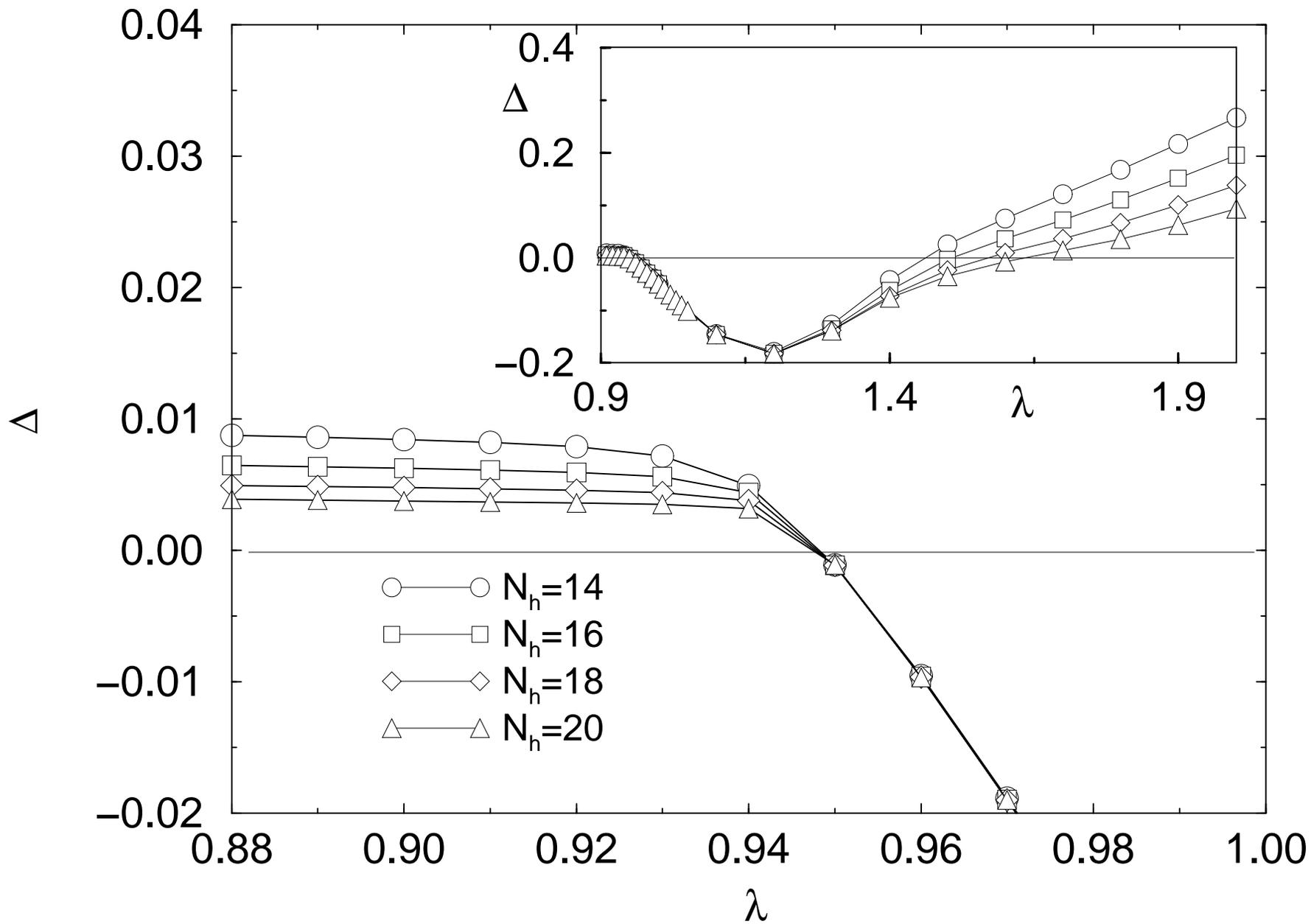
\begin{figure}
\caption[] {First excited state binding energy 
$\Delta = E_1 - E_0 - \omega$ as a function of $\lambda$. Results
are for $\omega=0.5$ and various Hilbert space sizes $N_h$.
Inset:  binding energy over a wider range of $\lambda$.
} 
\label{de}
\end{figure}

We begin by computing the energy difference $\Delta E = E_1
- E_0$, where $E_1$ and $E_0$ are the first excited state and the
ground state energies at $k=0$ (the two lowest
bands in Fig.~(\ref{spaghetti}) ).
In the case where the first excited
state of a polaron can be described as a polaron ground state
and an unbound extra phonon excitation, this energy
difference should in the thermodynamic limit equal the phonon
frequency, $\Delta E = \omega$.  In Fig.~(\ref{de}) we
plot the binding energy $\Delta = \Delta E -\omega $ for $\omega=0.5$
as a function of the electron-phonon coupling $\lambda$ for various 
sizes of the variational space.
We see two distinct regimes. Below $\lambda_c\sim 0.95$, $\Delta$ 
varies with the system size but remains positive $( \Delta >0 )$.  
Physically, for $\lambda < \lambda _c$, the additional phonon excitation
would prefer to be infinitely separated from the polaron,
but is confined to a distance no greater than $N_h -1$
by the variational Hilbert space.
As the system size increases, $\Delta $ slowly approaches zero from above
as the ``particle in a box'' confinement energy decreases.
In the other regime, $\lambda>\lambda_c$, our data has clearly converged
and $\Delta <0$.  This is the regime where the extra phonon excitation
is absorbed by the polaron forming an excited bound polaron.  
Since the excited polaron forms an exponentially decaying 
bound state, the method
already converges at $N_h=14$.  In the inset of Figure (\ref{de}) we
show the binding energy $\Delta$ in a larger interval of electron-phonon
coupling $\lambda$.  Although the results cease to converge at larger
$\lambda$, we notice that the binding energy $\Delta$ reaches a minimum as a
function of $\lambda$.  As we will demonstrate within the strong coupling
approximation, the true binding energy approaches zero exponentially from
below with increasing lambda.  In Figure~(\ref{phd}) we show the phase
diagram, valid for $k=0$, separating the two regimes.  The phase boundary, 
given by $\Delta=0$, was obtained 
numerically and using strong coupling perturbation theory in $t$ 
to first and second order. Details of the latter calculation are 
given in the next subsection.
The phase transition where $\Delta$ becomes negative at
sufficiently large $\lambda$ is also seen in exact
diagonalization calculations \cite{fehske1}.

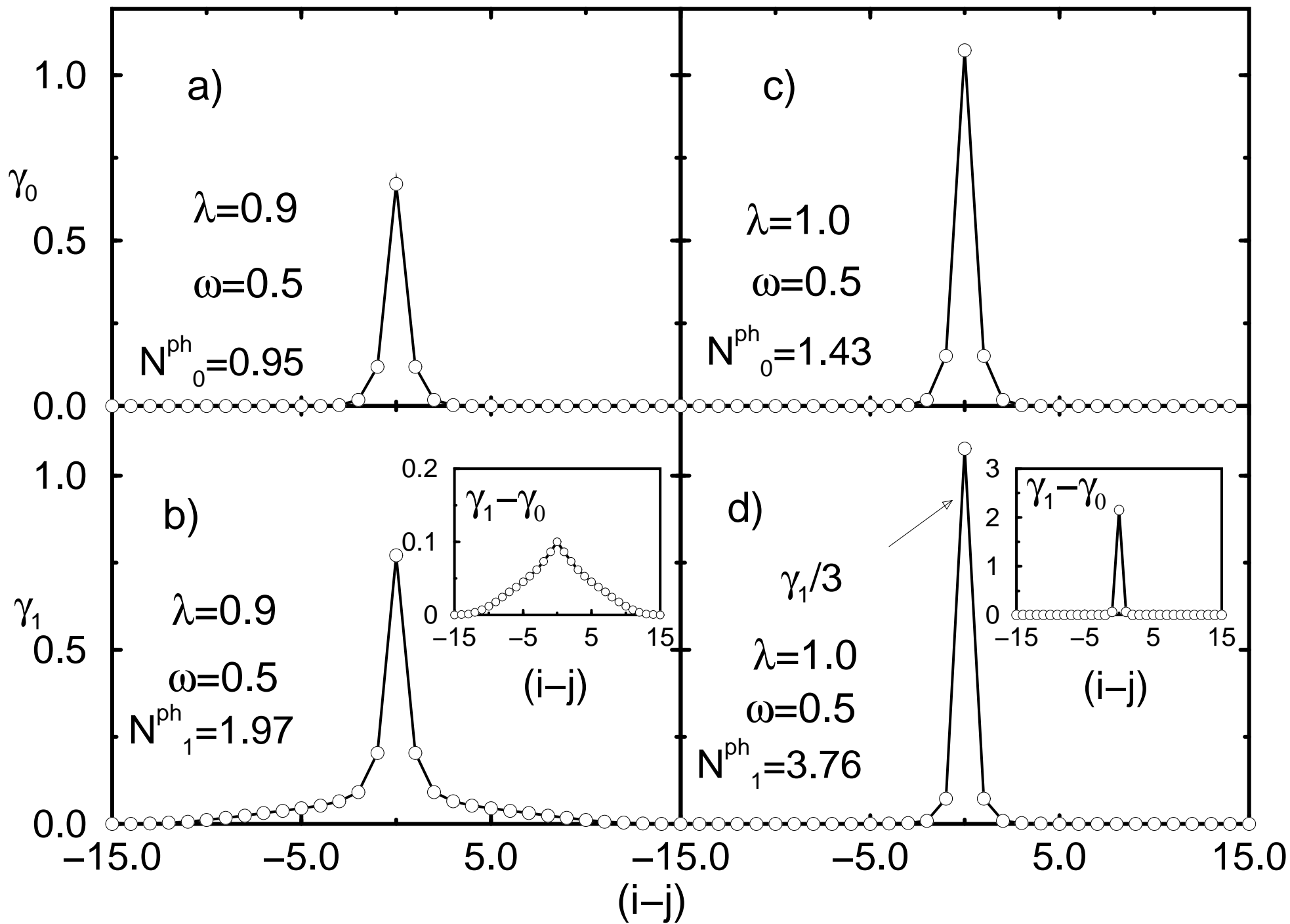
\begin{figure}
\caption[] {The phonon number $\gamma$ as a function of the distance
from the electron position $(i-j)$ for the ground state (a) and the
first excited state (b) both computed at $\lambda=0.9$ and the same in
(c) and (d) for $\lambda = 1.0$. All data are computed at 
phonon frequency $\omega=0.5$ and $N_h=18$.  
Note that (d) is a plot of $\gamma_1 / 3$.
Insets in (b) and (d)
represent differences $\gamma_1-\gamma_0$ as a function of the
distance $(i-j)$. In (b), $\gamma_1-\gamma_0$ drops to zero around
$\vert i-j\vert=15$. This is a finite-size effect. Computing the same
quantity with larger $N_h$ below the phase transition would lead
to a larger extent of the correlation function indicating 
that the extra phonon excitation is not bound to the polaron.  
}
\label{gammafig}
\end{figure}

In Figure (\ref{gammafig}) we compute
the distribution of the number of excited phonons in the
vicinity of the electron $\gamma(i-j)$, Eq.~(\ref{gamma}), for the
ground state $\gamma_0$ and the first excited state $\gamma_1$
slightly below $(\lambda=0.9)$, and above $(\lambda=1.0)$  
the transition for
$\omega=0.5$. The central peak of the correlation function $\gamma_1$
below the transition point is
comparable in magnitude to $\gamma_0$  (Figs.~(\ref{gammafig}a,b)). 
The main difference between the two curves is the long range
decay of $\gamma_1$ as a
function of distance from the electron, onto which the central peak is
superimposed. 
The extra phonon that is represented by this long-range
tail extends throughout the whole system and is not bound to
the polaron. See also the  difference $\gamma_1-\gamma_0$ 
in the inset of Fig.~(\ref{gammafig}b).
The existence of an unbound, free phonon is confirmed
by computing the difference of total phonon number $N^{ph}_{0,1}=\sum_l
\gamma_{0,1}(l)$. This difference should equal one below the
transition point. Our numerical values give $N^{ph}_1-N^{ph}_0\sim
1.02$.  We attribute the deviation from the exact result to
finite-size effects.

Correlation functions above the transition point (Figs.~(\ref{gammafig}c,d)) 
are physically different.  First, phonon correlations in $\gamma_1$ decay
exponentially, which also explains why the
convergence in this region is excellent.  Second, the size of the
central peak in $\gamma_1$ is 3 times higher than $\gamma_0$.
(Note that to match scales in Figure
(\ref{gammafig}d) we divided $\gamma_1$ by 3).
The difference in total phonon number gives
$N^{ph}_1-N^{ph}_0\sim 2.33$.  We are thus facing a totally different
physical picture: the excited state is composed of a polaron which
contains several extra phonon excitations (in comparison to the ground
state polaron) and the binding energy of the excited
polaron is $\Delta <0$. The extra phonon excitations are located
almost entirely  on the electron site (see the inset of 
Fig.~(\ref{gammafig}d)). The value of $\gamma_1-\gamma_0$ at $j=0$ is 2.16,
which almost exhausts the phonon sum.

\subsection{Strong-coupling perturbation theory}

In this section, we calculate the ground and
excited state energies of the polaron perturbatively
in the hopping $t$.  For $t=0$, the Hamiltonian
Eq.~(\ref{ham}) describes a harmonic oscillator
with a shifted origin (due to the $\lambda$ force term)
on the site with the electron,
and unshifted oscillators on the other sites.
The Hamiltonian in the new basis is given by
the canonical transformation $\tilde H = e^S H e^{-S}$, where
\begin{equation}
S = -g\sum_j n_j(a_j-a_j^\dagger),
\label{S}
\end{equation}
and $g=\lambda/\omega ~$ \cite{lang}. 
After some algebra (see also Ref. \cite{marsiglio}), the
transformed Hamiltonian takes the following form:
\begin{eqnarray}
\tilde H &=& H_0 + V \nonumber \\
H_0&=&\omega\sum_j  
a_j^\dagger a_j - \omega g^2 \sum_j n_j \nonumber \\
V &=& -t e^{-g^2}\sum_j \left(c_j^\dagger c_{j+1} 
e^{ g\left(a_j^\dagger - a_{j+1}^\dagger\right)} 
e^{-g\left(a_j - a_{j+1}\right)} + {\rm h.c.}\right) ~,
\label{hamtilde}
\end{eqnarray}
where $n_j = c_j^\dagger c_j$. 
The operator $a_j^\dagger$ in Eq.~(\ref{hamtilde})
creates a phonon excitation on site $j$ relative
to the shifted oscillator if there is an electron
on site $j$, and relative to an unshifted oscillator
on the other sites.  
The first term in $H_0$ is the energy of the phonon excitations,
and the second is the energy gained by the oscillator
that is displaced by the force of the electron.
In strong coupling perturbation
theory, $V$ in Eq.~(\ref{hamtilde}) is considered a
perturbation.  It represents the hopping of an electron,
including possible creation and destruction of
phonon excitations.

The lowest energy eigenstates of the unperturbed
Hamiltonian $H_0$ have no extra phonon excitations,
and an energy $E^{(0)}_0 = - \lambda ^2 / \omega $.
They can be written 
$\vert \phi_0(j)\rangle = c^\dagger_j\vert \cal O \rangle$, where
$\vert\cal O\rangle$ represents vacuum for electron and phonon degrees
of freedom. This state represents a polaron localized on the site
$j$. Evidently the ground state is $N-$fold degenerate, where $N$ is
the number of sites in the system.  The perturbation lifts this
degeneracy. The matrix elements 
$V_0(i,j)=\langle\phi_0(i)\vert V\vert\phi_0(j)\rangle$ 
can be readily
computed since the exponential factors in $V$ are not effective in this case,
\begin{equation}
V_0(i,j) = -t e^{-g^2};\ \ j = i \pm 1.
\label{v0}
\end{equation}
This describes a translation-invariant tight-binding model
in one dimension with nearest neighbor hopping.
To first order in $t$, the ground state energy
of a polaron with momentum $k$ is
\begin{equation}
E_0(k)=-\lambda^2/\omega - 2te^{-g^2}\cos k.
\label{e0}
\end{equation}

{\bf Excited states:}  The lowest energy states of
$H_0$ are the $N$ degenerate states of energy
$- \lambda ^2 / \omega $, considered above.
The next lowest energy sector contains the $N^2$ degenerate states
of energy $- \lambda ^2 / \omega ~+~ \omega$, of the form 
$\vert \psi_1(j,l)\rangle = c^\dagger_j a^\dagger_l\vert \cal O
\rangle$.  The electron is on site $j$ and the
additional phonon excitation is on site $l$.
We do degenerate perturbation theory to $O(t)$ in this 
excited sector.  Using the total momentum $k$ as a good
quantum number, the 2d tight-binding problem in $(j,l)$ becomes
a 1d tight-binding problem.  The 1d basis functions
are $\vert \phi_1(j)\rangle  = \vert \psi_1(0,j)\rangle$,
where $j$ is the distance between the phonon and the electron.
The nonzero matrix elements 
$V_1(i,j)=\langle\phi_1(i)\vert V\vert\phi_1(j)\rangle$ are
\begin{eqnarray}
V_1(0,0) &=& -2t g^2 e^{-g^2}\cos k \nonumber \\
V_1(0,\pm1) &=& -t (1-g^2) e^{-g^2} \nonumber \\
V_1(-1,1) &=& -t g^2 e^{-g^2} e^{ i k} \nonumber \\
V_1(i,j) &=& -te^{-g^2};\ \ j=i\pm 1; i,j \neq 0.
\label{v1}
\end{eqnarray}
This is a 1d tight-binding model that is translation-invariant
except near the origin, where there is a second-neighbor hopping term
and other modifications.  
The matrix $V_1(i,j)$ is Hermitian.  
Matrix elements in Eq.~(\ref{v1}) define a secular equation
$\vert V(k)-E(k)\vert =0$ 
for the energies that can easily be solved numerically for a large system.
For each total wavevector $k$, there are $N$ independent solutions.
The lowest energy first excited state at momentum $k$ is
\begin{equation}
E_1(k)=-\lambda^2/\omega + \omega + E_1^{(1)}(k),
\label{e1}
\end{equation}
where $ E_1^{(1)}(k)$ is the lowest energy solution of the secular equation.

The numerical and analytic solution of the secular equation
reveals that there is a true phase transition 
(energy nonanalytic in the parameters) for the first
excited state.  This is perhaps surprising in light
of the theorem that there can be no phase transition
for the ground state \cite{gerlach}, and demonstrates the impossibility
of extending the theorem to include excited states.
We consider first a total momentum
$k=0$.  For $g > g_1 = 1$, the lowest energy first excited state
is found to be a bound state of a polaron and an additional phonon.
The bound state is raman active.  
For $g < g_1$, the first excited state is unbound,
with an energy exactly $\omega$ higher than the ground state.
This energy is in fact an upper bound for the first
excited state in any dimension and at any $\vec k$,
since one can construct a variational state with
a zero momentum polaron and a momentum $\vec k$
phonon at infinite separation.
The energy of the first excited
state is nonanalytic (discontinuous first derivative)
at $g=g_1$.
The bound state formation is unusual for several
(related) reasons:  (a)  For $g > g_1$, the binding
energy is linear in $(g-g_1)$ for small $(g-g_1)$, 
rather than the $(g-g_1)^2$
dependence that is typical for 1d bound state problems.
(b)  For $g < g_1$, the phase shift is zero
between the polaron and an additional unbound phonon 
of relative momentum $q \rightarrow 0$.
The polaron and additional phonon pass through each
other transparently,
in contrast to the usual repulsive phase shift.
(c)  At $g=g_1$, even for finite $N$ (periodic boundary conditions),
there are two exactly degenerate ground states, in contrast
to the usual avoided crossing.
These unexpected properties result in part from the fact
that the central site in the tight-binding model (Eq.~\ref{v1})
becomes uncoupled from the rest of the lattice precisely
at $g=g_1$.
The binding energy in 1d can in fact be determined analytically
from Eq.~(\ref{v1}).  One can show that for
$g \geq g_1$, $E^{(1)}_1 (k) = -t e ^ {-g^2} (x^2+1)/x$, where
\begin{eqnarray}
x &=& { 3 g^2 - 1 + \sqrt { (9 g^2-1) (g^2-1) }  \over 2} ~~;~~ k=0  
\nonumber \\
x &=& g^2 ~~;~~ k= \pi ~.
\label{analytic}
\end{eqnarray}

The $O(t)$ strong-coupling analysis can be extended to
higher dimensions and to nonzero total momentum $\vec k$.
The $\vec k = 0$ state always has an energy less than
or equal to that at any other $\vec k$.
We find that for $g < g_1$, the first excited state
at any momentum $\vec k$ is unbound, with $g_1 = 1$
in any dimension $d \ge 1$.  The $k=0$ phase shift
also vanishes for $g < g_1$ in any dimension.
For $g > g_2$, the first excited state
at any momentum is bound \cite{bound}.  For $g_1 < g < g_2$,
there is a phase transition on a surface
in $\vec k$-space such that inside the surface
(including $\vec k=0$) 
the first excited state is bound,
and outside the surface (including $\vec k= ( \pi, \pi, \dots )~)$
the first excited state is unbound.  The location of the
surface in $\vec k$-space is $g$-dependent.  The numerically obtained
values are $g_2 = 2.1$ in 3d,
$g_2 = 1.66$ in 2d, and $g_2 = g_1 = 1$ in 1d.
(1d is special in that there is no intermediate phase.
However, as $g \rightarrow g_1 ^+$ in 1d, the $k=\pi$ state
is very weakly bound compared to $k=0$; see Eq.~(\ref{analytic}).
The $k=\pi$ state has zero amplitude 
to have the electron and the additional phonon on the same site.)
The binding energy at $\vec k = 0$
is linear in $(g-g_1)$ for higher dimensions,
as it is in 1d.

Strong coupling perturbation theory is carried
to $O( t^2 )$ in the appendix.

\section{Conclusions}

In summary, we have presented a variational approach for solving the
Holstein model with dynamical, quantum phonons
based on an exact diagonalization method. The
variational space is defined on an infinite lattice. It is constructed
by successive application of the off diagonal terms of the Hamiltonian
starting from a single electron state.  This leads to a 
systematically improvable variational
basis that turns out to be efficient for calculating the
ground state and low-lying excited state properties of the model. 
The method can compute properties in all parameter regimes,
but it is at its best in the intermediate coupling regime,
where strong and weak coupling perturbation theories, 
and other variational methods have problems.

The method allows the computation of energy bands and other physical
properties at continuous wavevectors. For intermediate coupling
strengths we are able to reach an accuracy of 12 digits for 
the ground state energy at small $k$
with as few as $M=88052$ basis states and only a few seconds of CPU
time on a workstation.  Our results for energy bands presented in
Fig.~(\ref{ekfig}) compare well with other numerically more intensive
methods and are more precise at both large and small $k$ than some of them.  
Our energies are variational in the thermodynamic limit for any $k$.
We believe that all results shown for energy bands computed
with $M=3 \times 10^6~$ states converge to at least 4 digits for
arbitrary $k$.

The accuracy of our method can be seen from the comparison of
effective masses in Fig.~(\ref{mass0}). While our results have
converged to at least 4 digits in all parameter regimes presented
in Fig.~(\ref{mass0}), GL and DMRG methods give less reliable results.
While deviations of DMRG are insignificant, deviations of GL results
near strong coupling seem to be systematic.  Results for the
quasiparticle residue as a function of the wavevector 
in the weak coupling regime show
a smooth crossover between the predominantly
electronic character to a predominantly phononic character of the
polaron. Our results agree with previous ED calculations. 

The correlation functions $\chi$ agree well with DMRG results in the
intermediate coupling regime. 
In the weak coupling regime our method gives more
reliable results.
Close to the extreme strong coupling and
adiabatic regime, our correlation functions
have not fully converged as a function of $N_h$
for $N_h=21$, $M=6 \times 10^6$.  Our results
in this regime are qualitatively similar to, but
less accurate than those of the DMRG.
Correlation functions $\chi$ computed at
different $k$ provide detailed information on how the 
weak coupling polaron
transforms as $k$ increases from
$k=0$ to $k=\pi$. 

Using numerical and strong-coupling approaches,
we find a true phase transition (rather than a crossover)
in the first excited state, where a polaron plus phonon
system changes from unbound at weak coupling to bound
at strong coupling.  
The first excited state does not contribute to the
optical conductivity, but rather is Raman active.

There are a number of extensions to this work
that we have not included in this paper.
It is straightforward to consider anharmonic phonons.
(In fact, the extreme double-well limit where only
the lowest two states are retained is numerically
less demanding than the linear case.)
The AC conductivity and spectral function of
a polaron can be calculated by the same methods.
Properties of other Hamiltonians, including those with SSH-type couplings
where phonons modify the hopping $t$ can be calculated.
Extensions that allow certain phonon excitations
to be infinitely far from the electron are possible.
Properties in higher 
spatial dimensions can be calculated.
One can also calculate the properties of bipolarons,
including Hubbard onsite and longer range interactions
between electrons.  And finally, one can calculate
the coherent quantum dynamics of electron-phonon
coupled systems driven far from equilibrium using 
similar methods \cite{janez1,janez2}.

\bigskip\bigskip

We would like to acknowledge valuable conversations
with A.~Alexandrov, A.~Bishop, H.~Fehske, E.~Jeckelmann, L.~C.~Ku,
K.~Lindenberg, S.~Marianer, F.~Marsiglio, H.~R\" oder, and S.~White.
J.B. would like
to acknowledge the hospitality of Los Alamos National Laboratory where
the major part of this work was performed.
This work was supported by the US Department of Energy.

\appendix
\section*{}

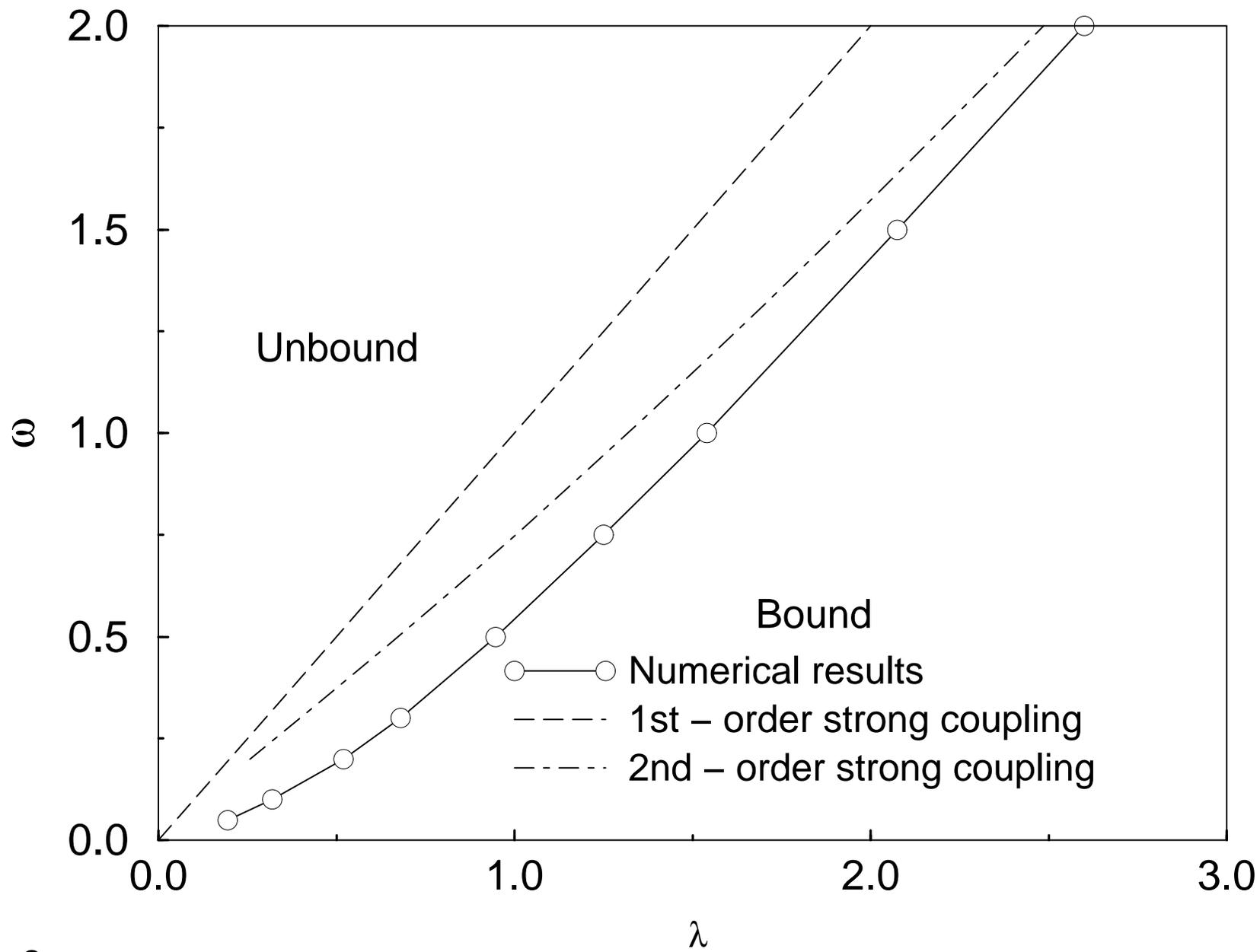
\begin{figure}
\caption[] {The phase diagram for the bound to unbound
transition of the
first excited state, obtained using 
the condition $\Delta (\omega,\lambda)=0$. 
The corresponding phase diagram for the ground state would be
blank---there is no phase transition
in the ground state, only a crossover.
} 
\label{phd}
\end{figure}

For simplicity we have limited our calculation of the energy
corrections in second order strong-coupling perturbation theory
to 1d and zero momentum, $k=0$. Following the work of Marsiglio
\cite{marsiglio}, the ground state correction to second order in
the hopping $t$ is given by
\begin{equation}
E_0^{(2)} = -2t^2{e^{-2g^2}\over \omega}\left [
\sum_{n,m = 1}{g^{2(m+n)}\over n!m!}{1\over n+m} +
3\sum_{n = 1}{g^{2n}\over n!}{1\over n} \right].
\label{e02}
\end{equation}
Calculation of the energy corrections of the excited state energy
$E_1$ involves degenerate perturbation theory, where matrix elements
between degenerate states $\vert \phi_1(j)\rangle$ are computed to
second order in $t$. After a straightforward but tedious
calculation we obtain for the non-zero matrix elements
\begin{eqnarray}
V_1(0,0) = &-&2tg^2 e^{-g^2} \nonumber \\ 
           &+& 2t^2{g^2e^{-2g^2}\over \omega}\left [
\sum_{n,m = 1}{g^{2(m+n)}\over n!m!}{(1-{n\over g^2})^2\over 1-n-m} +
\sum_{n = 2}{g^{2n}\over n!}{(1-{n\over g^2})^2+2\over 1-n} +2 \right]
\end{eqnarray}

\begin{eqnarray}
V_1(0,1) = &-&t(1-g^2) e^{-g^2}\nonumber \\ 
           &-& t^2{g^2e^{-2g^2}\over \omega}\left [
\sum_{n,m = 1}{g^{2(m+n)}\over n!m!}{(1-{n\over g^2})(1-{m\over g^2})
\over 1-n-m} +
3\sum_{n = 2}{g^{2n}\over n!}{(1-{n\over g^2})\over 1-n} +2 \right]
\end{eqnarray}

\begin{eqnarray}
V_1(1,1) = t^2{e^{-2g^2}\over \omega}&{\Bigg [}&
\sum_{n,m = 1}{g^{2(m+n)}\over n!m!}\left({g^2(1-{n\over g^2})^2
\over 1-n-m} -{1\over n+m}\right) \nonumber \\
&+&\sum_{n = 2}{g^{2n}\over n!}{2g^2-2n+{n\over g^2}\left 
(n+2\right )\over 1-n} +g^2 {\Bigg ]}
\end{eqnarray}

\begin{equation}
V_1(-1,1) = -tg^2 e^{-g^2} + t^2{g^2e^{-2g^2}\over \omega}\left [
\sum_{n = 2}{g^{2n}\over n!}{(1-{n\over g^2})^2\over 1-n} 
+1 \right]
\end{equation}

\begin{equation}
V_1(0,2) =  -t^2(1-g^2){e^{-2g^2}\over \omega}
\sum_{n = 1}{g^{2n}\over n!}{1\over n} 
\end{equation}

\begin{equation}
V_1(j,j+2) =  -t^2{e^{-2g^2}\over \omega}
\sum_{n = 1}{g^{2n}\over n!}{1\over n} ; \ \ \ j\geq 1
\end{equation}

\begin{equation}
V_1(j,j) = -2t^2{e^{-2g^2}\over \omega}\left [
\sum_{n,m = 1}{g^{2(m+n)}\over n!m!}{1\over n+m} +
2\sum_{n = 1}{g^{2n}\over n!}{1\over n} \right];\ \ \  j\geq 2
\end{equation}

\begin{equation}
V_1(j,j+1) =  -t e^{-g^2} ;\ \ \ \  j\geq 1 .
\end{equation}
For $k=0$, $V_1(i,j) = V_1(j,i) = V_1(-i,-j)$.
We numerically solve
the secular equation $\vert V - E\vert =0$. The lowest eigenvalue
$E_1^{(1,2)}$ of the secular equation gives us corrections to the
excited state energy $E_1$ to second order in the hopping $t$,
\begin{equation}
E_1=-\lambda^2/\omega + \omega + E_1^{(1,2)}.
\label{e12}
\end{equation}
The binding energy to second order at $k=0$ is given by 
Eqs.~(\ref{e0}), (\ref{e02}), and (\ref{e12}),
\begin{equation}
\Delta = E_1^{(1,2)}+2te^{-g^2}-E_0^{(2)}.
\label{d2}
\end{equation}
The second order results are used to calculate
the phase boundary in Fig.~(\ref{phd}).

\begin{figure}
\caption[]{ The ground and excited state energy eigenvalues
(those $E_j \leq 0$) are plotted 
as a function of $k$ (in units of $\pi$)
for $\lambda = \omega = 1$, $N_h = 9$, $M=1185$.
Excited states consist of the polaron with additional
bound or unbound phonon excitations.
The hopping parameter $t=1$ is assumed here and throughout
this work wherever $t$ is not explicitly specified.
}
\label{spaghetti}
\end{figure}

\begin{figure}
\caption[] {Lattice deformation $\chi$ as a function  of $(i-j)$ for
(a) $\omega =1.0$ and $\lambda=0.5$,
(b) $\omega =0.1$ and $\lambda=0.1$, and
(c) $\omega =0.1$ and $\lambda=0.435$. Our results are represented by 
filled symbols and those of Jeckelmann and White \cite{white} 
by open circles. In case c) 
our results for  $N_h = 21$ have not yet reached
the large $N_h$ limit.
The star represents an extrapolation of our data (assuming
exponential convergence) to
$N_h\to \infty$, $\chi (0) = 5.5 \pm 0.1$.
 } 
\label{chi}
\end{figure}

\begin{table}
\caption{} 
\begin{tabular}{|c|cccc|}
$\lambda /\omega$ &   Present       &  ED N=16    &  DMRG N=32 &  Global-Local \\ \hline 
     1            & -2.469684723933 & -2.46968477 & -2.46968   & -2.46931      \\
$\sqrt{2}$        & -2.998828186867 & -2.99882816 & -2.99883   & -2.99802      \\
\end{tabular}
\label{energ}
\end{table}


\begin{references}
\bibitem{trade} It is possible to construct a variational space
with a different tradeoff between phonon number and phonon distance.
It is sometimes useful at very strong coupling to add
to the Hilbert space states with a very large number of
phonon quanta on the electron site.
\bibitem{bound}  The first excited state at nonzero
momentum $\vec k$ is a bound state
if  $E_1(\vec k) - E_0(\vec k = 0) < \omega$,
so that decay into a polaron and a phonon is energetically forbidden.
\end{references}
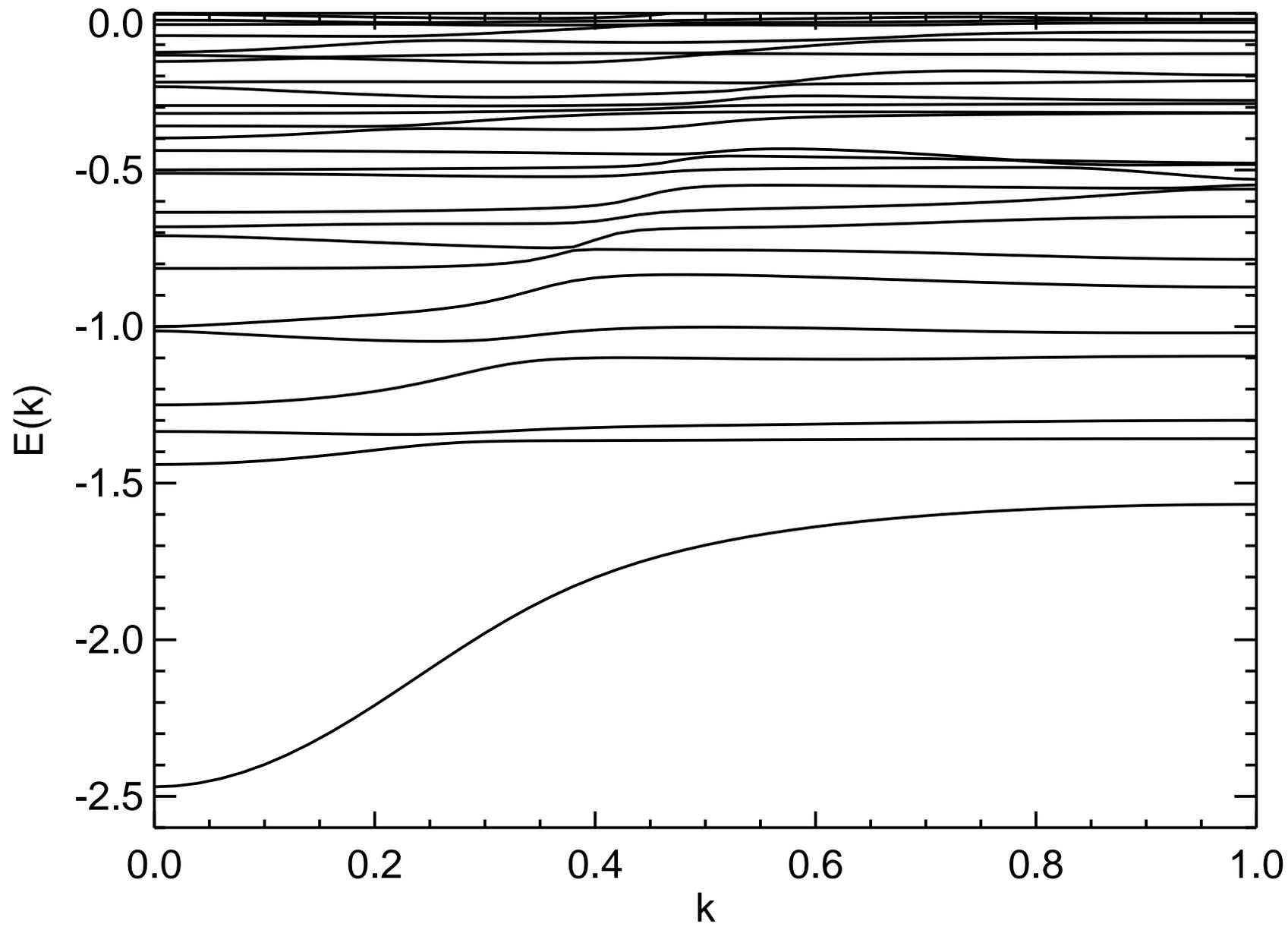
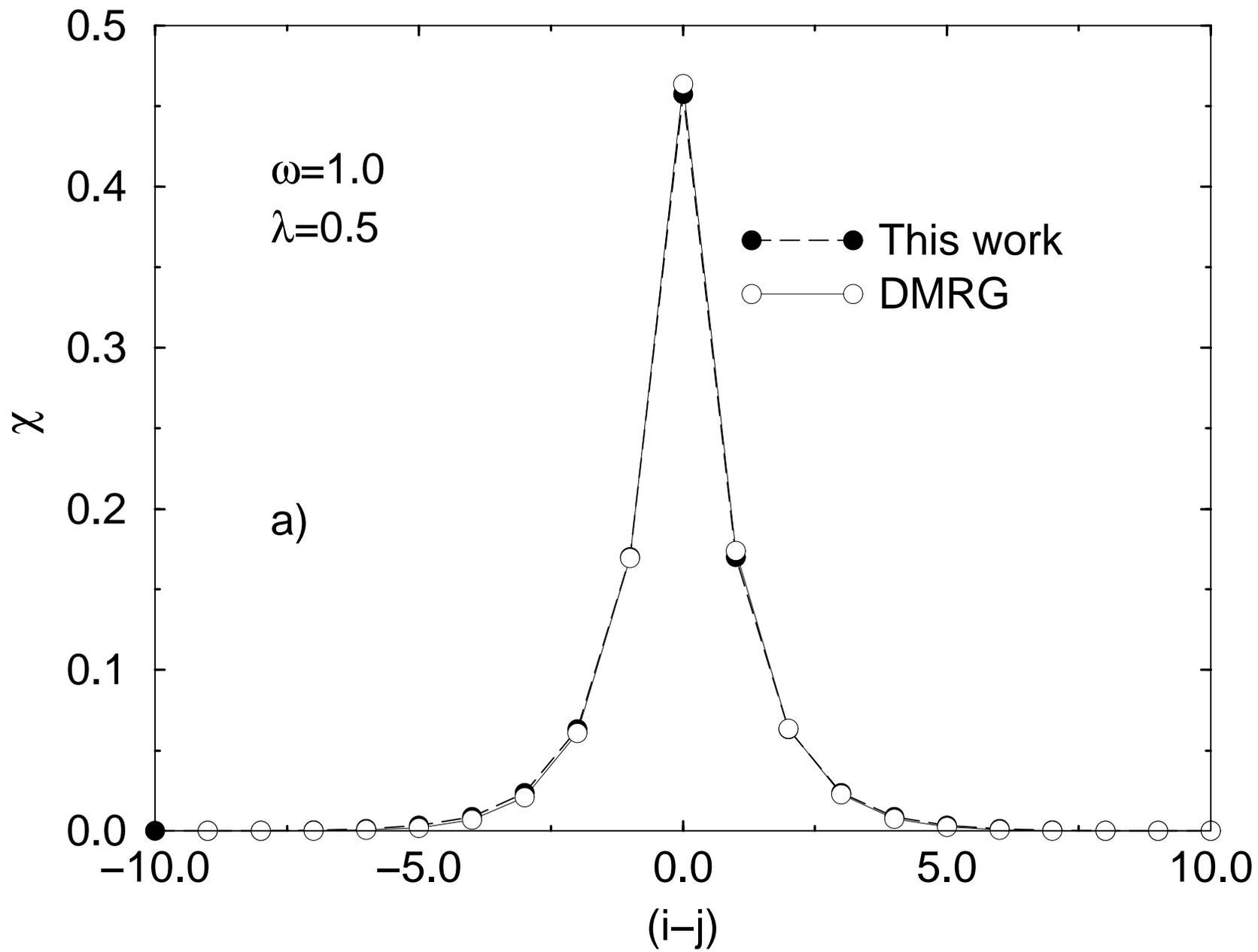
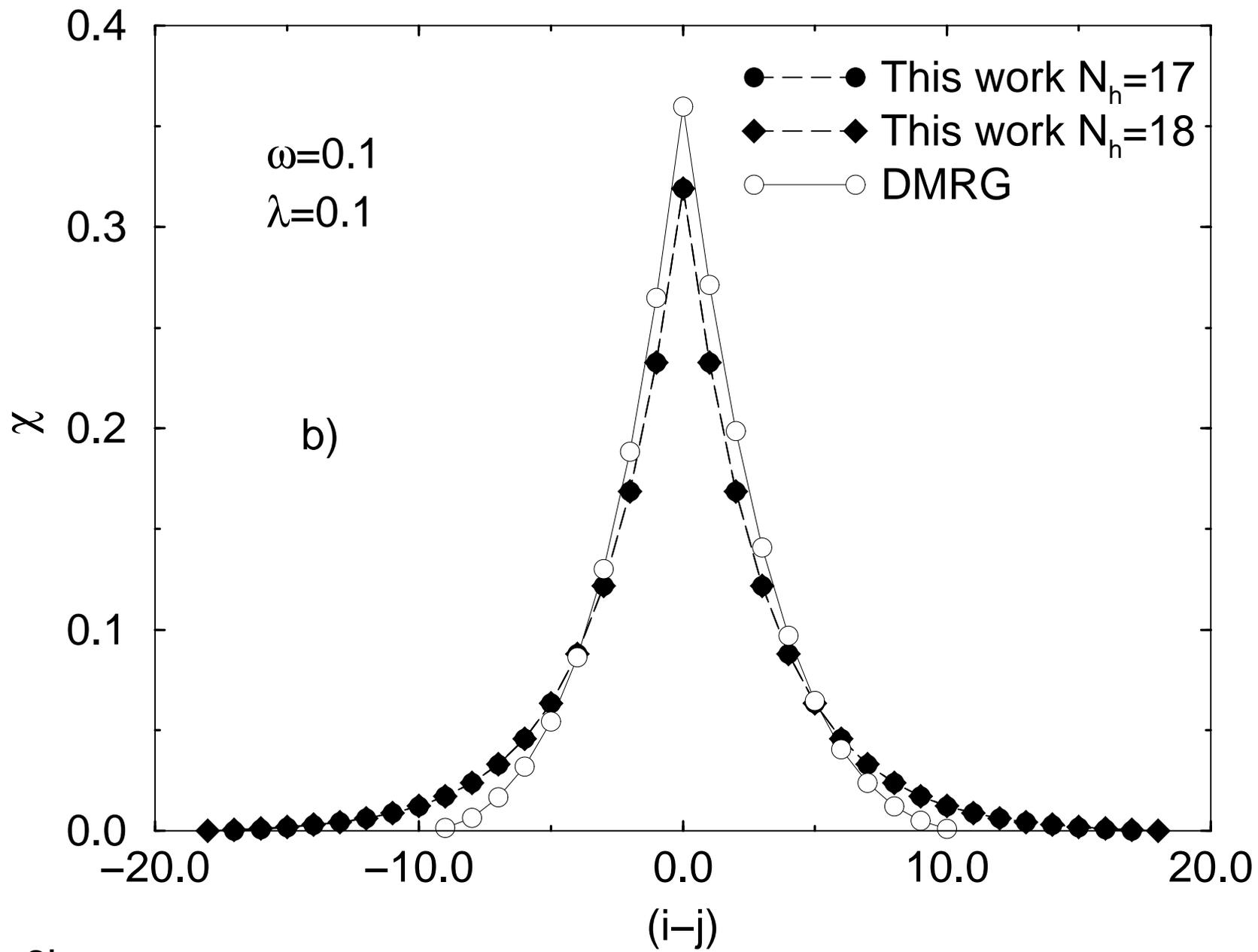
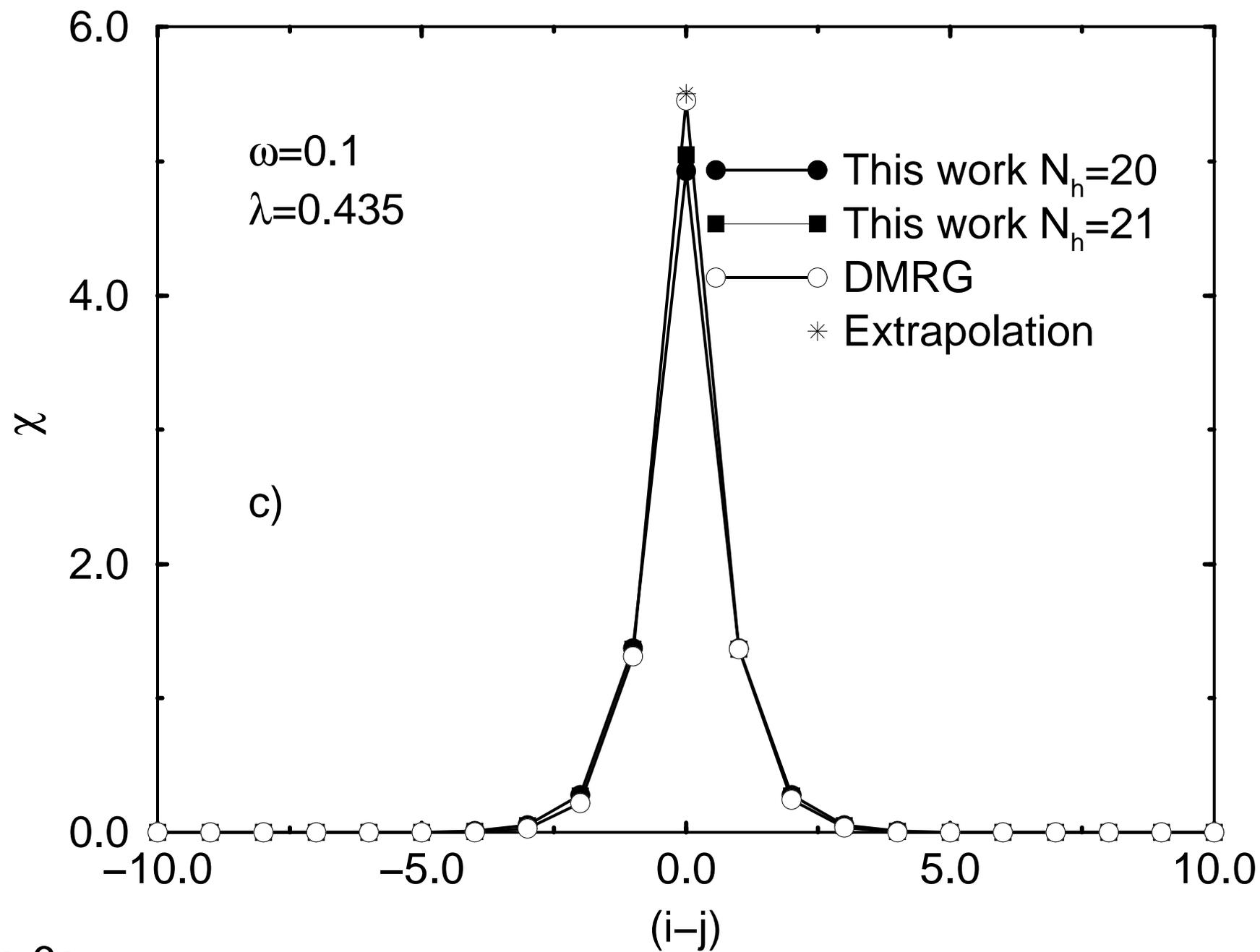
\end{document}